\documentclass[prb,twocolumn]{revtex4}

\usepackage{psfrag}
\usepackage[dvips]{graphicx}
\graphicspath{{./}{Figs/}}

\usepackage{amsmath}
\usepackage{amssymb}
\usepackage[english]{babel}

\usepackage[latin1]{inputenc}
\usepackage{lmodern}
\usepackage[T1]{fontenc}

\DeclareMathOperator{\e}{{\displaystyle e}}

\newcommand{\pp}{\cfrac{1}{\pi} P \int_{-\infty}^{+\infty} }
\newcommand{\de}{\mathrm{d}}

\begin{document}

\title{Some integrals related to the Fermi function}
\author{G. Bevilacqua}
\affiliation{Department  of Information  Engineering  and Mathematical
  Sciences, University of Siena, Via Roma 56, I-53100 Siena, Italy}


\begin{abstract}
  Some elaborations  regarding the  Hilbert and Fourier  transforms of
  Fermi  function  are presented.   The  main  result  shows that  the
  Hilbert transform  of the difference  of two Fermi functions  has an
  analytical  expression in  terms of  the $\Psi$  (digamma) function,
  while  its Fourier  transform  is expressed  by  mean of  elementary
  functions.   Moreover  an  integral  involving the  product  of  the
  difference  of two  Fermi functions  with its  Hilbert  transform is
  evaluated   analytically.   These   findings   are  of   fundamental
  importance  in  discussing the  transport  properties of  electronic
  systems.
\end{abstract}

\maketitle

\section{Introduction}
\label{sec:intro}

In many problem in electronic transport calculation the Hilbert
transform  (HT)  of the  difference  of  two  Fermi function  (FF)  is
required\cite{appl:prb}.   Usually  this  task  is accomplished  in  a
numerical way using the Matsubara expansion of FF or other approximate
expansions             with             faster             convergence
behaviour\cite{ff:expa,ff:expa:2,ff:expa:3:cf, ff:expa:4}.
While  these  methods as  well  as  the  Sommerfeld expansion  are  of
valuable help in  numerical work, the importance of  exact results can
be hardy overlooked. 
 
In this paper I will show that the Hilbert and Fourier transform of
the difference of two FF can  be expressed in a closed analytical form
by means of the digamma function.  Moreover an universal integral
needed in the theory of electronic transport in presence of phonon
scattering\cite{appl:prb} is exactly evaluated.

The  paper  is organized  as  follow:  in  Section \ref{sec:hilb_t}  a
quickly  description of the  contour integration  details used  in the
work is  given; in Section  \ref{sec:HT:FF} the main  results are
derived,  while  in Section  \ref{sec:int:ggH}  an universal  integral
needed in the discussion of the asymmetric contribution to the current
flowing in  a molecular wire  is worked out.   Section \ref{sec:concl}
contains the conclusions.

\section{Hilbert transform}
\label{sec:hilb_t}
Given a function $g(x)$ its HT is defined as
\begin{equation}
  \label{eq:ht:def}
  {\cal H}(g) (y) \equiv g_H(y) \equiv \pp  \frac{g(x)}{y - x} \; \de x
\end{equation}
where $P$ is the usual Cauchy principal value. 

\begin{figure}[h]
  \centering
  { \psfrag{Imx}{$\Im (x)$}
    \psfrag{Rex}{$\Re (x)$}
    \psfrag{G}{$\Gamma$}
    \psfrag{A}{$A$}
    \psfrag{B}{$B$}
    \psfrag{C}{$C$}
    \psfrag{D}{$D$}
    \psfrag{y}{$y$}
    \psfrag{R}{$R$}
    \psfrag{r}{$r$}
    \psfrag{c}{$\gamma$}
    \psfrag{g}{$\gamma$}
  \includegraphics[scale=.50]{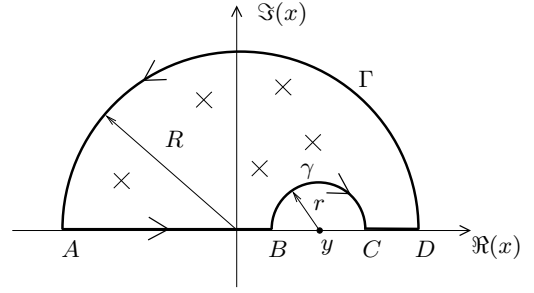} }
  \caption{Circuit in the complex plane used to evaluate Hilbert
    transform. The $\times $ symbols denote the poles of the
    integrand located at $x_n$ points.}
  \label{fig:circ}
\end{figure}

One powerful technique often used in the evaluation of such kind of
integrals extends the function involved in the complex plane and makes
use  the residua theorem. The typical situation is shown in Fig.~\ref{fig:circ}

By the residua theorem the integral over the closed circuit, done in
counter-clockwise way, is proportional to the sum of the internal
residua, while the integral over the real axis interval $[A,B]\cup
[C,D]$ 
in the limit of $R \rightarrow \infty$, and, separately 
$r \rightarrow 0$ gives the HT desired. More precisely
\begin{equation}
  \label{eq:circ}
  \begin{split}
  \int_{\mathrm{full \; circuit}} & = 2 \pi i \; \sum_n \; \mathrm{Res}(x_n) \\
                             & = \int_{\Gamma} + \int_{[A,B]\cup[C,D]} 
                                 + \int_{\gamma}
  \end{split}
\end{equation}

If the integral 
over the $\Gamma$ half-circumference vanish in the limit $R
\rightarrow \infty$ one has 
\begin{equation}
  \label{eq:circ:senza:G}
  2 \pi i \; \sum_n \; \mathrm{Res}(x_n) =  
  \int_{[-\infty,B]\cup[C,+\infty]} + \int_{\gamma}, 
\end{equation}
while in the limit $r \rightarrow 0$ the first integral in the
right-hand-side of \eqref{eq:circ:senza:G} becomes essentially the
principal value and the latter can be related to the residuum in
$y$, i.e.  $ \int_{\gamma} \rightarrow - \pi i \; \mathrm{Res}(y) $, where the minus 
sign is due to the clockwise orientation of the path. 

In summary one arrives at the following result
 \begin{equation}
  \label{eq:hilb:fin}
\pp  \frac{g(x)}{y - x} \; \de x = 
  2 \, i  \; \sum_n \; \mathrm{Res}(x_n) + i \; \mathrm{Res}(y)
\end{equation}
where $x_n $ are the poles of $g(x)$ in the $\Im(x) > 0$
half-plane. In case $g(x)$ has no poles on the real axis the preceding
formula simplifies as 
\begin{equation}
  \label{eq:hilb:fin:2}
  \pp  \frac{g(x)}{y - x} \; \de x = 
  2 \, i  \; \sum_n \; \mathrm{Res}(x_n) - i \; g(y).
\end{equation}

\section{Hilbert and Fourier transform of the difference of two Fermi functions}
\label{sec:HT:FF}
\begin{figure}[ht]
  \centering
 \centering
  \resizebox{0.45\textwidth}{!}{\includegraphics{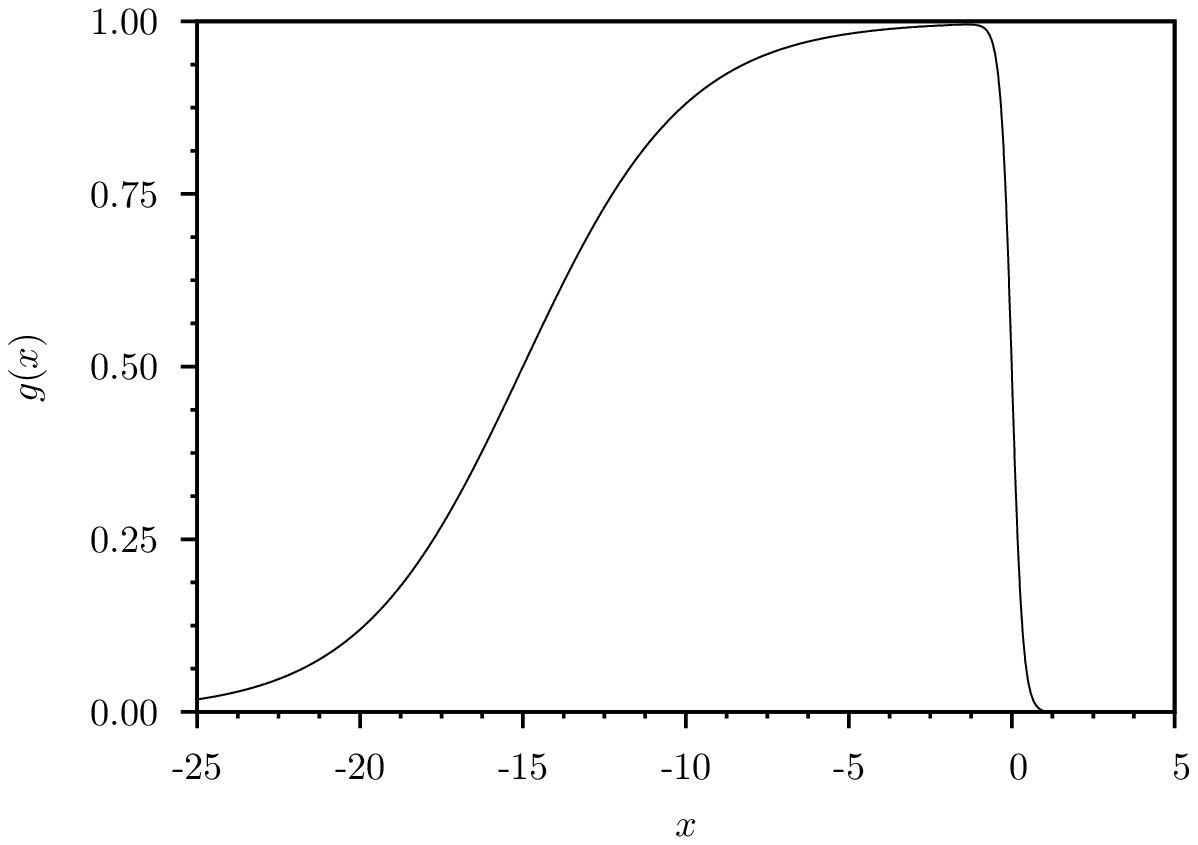}}\\
  \resizebox{0.45\textwidth}{!}{\includegraphics{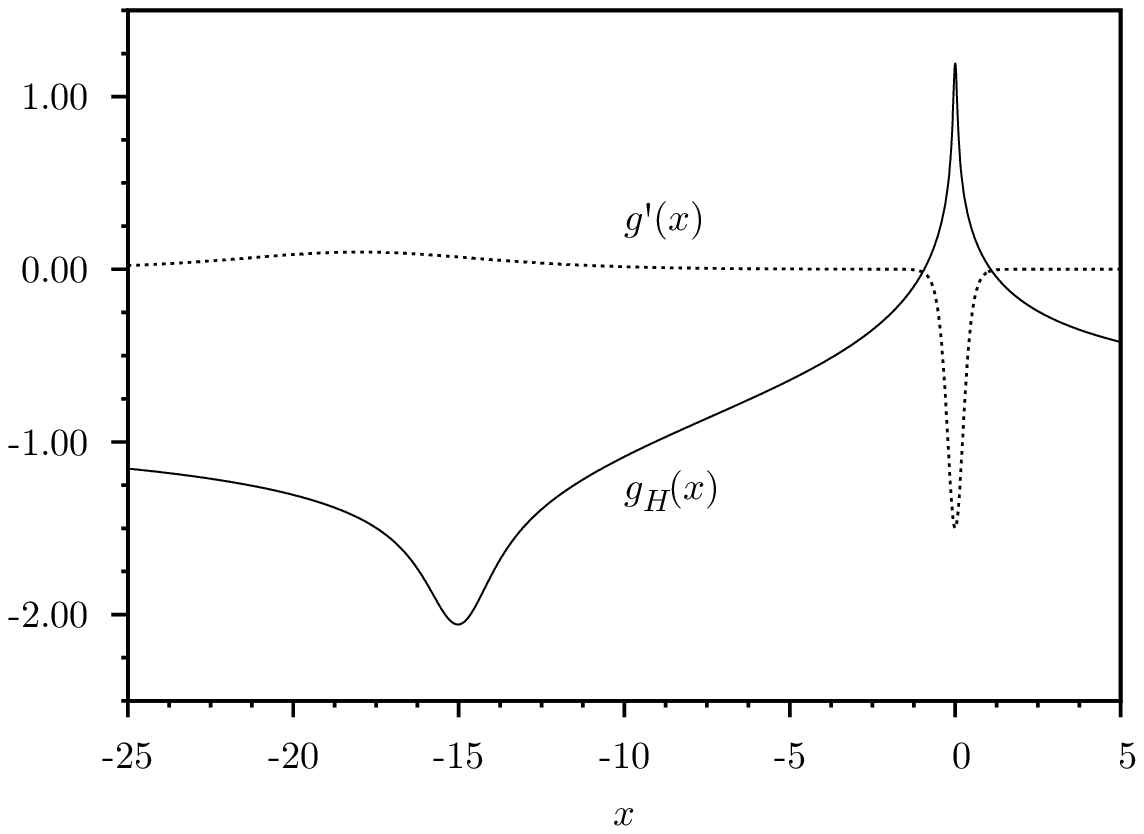}}
  \caption{(Top) Difference of two Fermi functions. (Bottom) Hilbert
    transform (solid line) and first derivative (dashed line) of the top
    panel. The parameters are $\mu_1 = 0$, $\mu_2 = -15$, $\beta_1 = 6$ and
  $\beta_2 = 0.4$.}
  \label{fig:FF_HT}
\end{figure}

Let apply the above machinery to the difference of FF
\begin{equation}
  \label{eq:ff}
  g(x)
  \equiv 
\cfrac{1}{\e^{\beta_1(x - \mu_1)} + 1} -
\cfrac{1}{\e^{\beta_2(x - \mu_2)} + 1} =
f_1(x) - f_2(x) 
\end{equation}
where  the  chemical   potentials  $\mu_i$  and  inverse  temperatures
$\beta_i$ can be thought as parameters.  The poles structure of $g$ is
composed of two infinite series i) $x_n = \mu_1 + (n + 1/2 )\; 2 \pi i
/\beta_1 $, $n = 0, \pm 1, \pm 2, \ldots $ with residua $-1/\beta_1$;
ii) $x_m = \mu_2  + (m +1/2 )\; 2 \pi i /\beta_2 $, 
  $m = 0, \pm 1, \pm 2, \ldots  $ 
  with residua
  $1/\beta_2$.
Only those with $n,m > 0$ are required. Collecting 
all the contributions one obtains
\begin{equation}
  \label{eq:FF:hilb}
  \begin{split}
    g_H(y)
    &  =   
    + 2 \,i \; \sum_{n=0}^{\infty} 
    \left[ 
      (x-x_n) \cfrac{f_1(x) }{y-x} - (x-x_n) \cfrac{f_2(x) }{y-x}
    \right]_{x=x_n} 
      \\ 
      & \quad  + 2\, i\; \sum_{m=0}^{\infty} 
      \left[ 
        (x-x_m) \cfrac{f_1(x) }{y-x} - (x-x_m) \cfrac{f_2(x) }{y-x}
      \right]_{x=x_m} \\
     & \quad  - i \left[ f_1(y) - f_2(y) \right].
  \end{split}
\end{equation}
Noticing  that  only the  first  and the  fourth  term  in the  square
brackets  in the  right-hand side  are different  from zero  and equal
respectively  to $  (-1/\beta_1)/(y-x_n)$  and $(-1/\beta_2)/(y-x_m)$,
after some straightforward manipulations 
one obtains
\begin{equation}
  \label{eq:FF:hilb:2}
  \begin{split}  
    g_H 
    = & 
    2 \,i \; \sum_{n=0}^{\infty} 
    \left[ \frac{-1/\beta_1}{y - \mu_1 - (n+1/2) {2 \pi i}/{\beta_1} }
    \right. \\
    & \phantom{ 2 \,i \; \sum_{n=0}^{\infty} 
      \left[ \right. } 
    \left. -\frac{-1/\beta_2}{y - \mu_2 - (n+1/2) 2 \pi i /\beta_2 } 
    \right] \\ 
    & - i \left[ f_1(y) - f_2(y) \right].
  \end{split}
\end{equation}
Introducing the  quantities $w_j= 1/2+ i  \beta_j\,(y-\mu_j)/2\pi $, the
sum can  be recast in  a form that  shows the relation with  the Euler
digamma function $\Psi$\cite{abramowitz:stegun,erdelyi}
\begin{equation}
 \label{eq:FF:hilb:sum}
 \sum_{n=0}^{\infty} 
    \left( \frac{1}{n + w_1 } 
      - \frac{1}{n + w_2}
    \right) 
    = 
      \Psi\left( w_2 
      \right)
      -
      \Psi\left( w_1
      \right)
\end{equation}
The term $f_1(y) - f_2(y)$ can be further elaborated exploiting the relation between the
$\Psi$ and the FF
\begin{equation}
  \label{eq:FF:Psi}
  f(z) = \frac{1}{2} 
  + \frac{1}{\pi} \Im \left[ \Psi \left( 1/2 - i\,z/2\pi\right) \right]
\end{equation}
%
so  after  some  manipulations   and  remembering  that  $\Psi(z^*)  =
\Psi(z)^*$ one 
finds the analytical result 
\begin{equation}
  \label{eq:FF:hilb:fin}
  g_H(y;\mu_1,\mu_2,\beta_1,\beta_2 ) = \frac{1}{\pi}
  \Re \left[  
    \Psi\left( w_2 \right) 
     - \Psi\left( w_1 \right) \right].
\end{equation}
In Fig.~\ref{fig:FF_HT} a particular case is shown.

Various properties of $g_H$ can be inferred by inspection. For
instance 
\begin{equation}
  \label{eq:gH:infi}
  g_H(y) = \frac{1}{\pi} \ln \frac{\beta_2}{\beta_1} 
  + \frac{\mu_1 - \mu_2}{y} + O(1/y^2)
\qquad
  |y| \rightarrow \infty
\end{equation}

Moreover if $\beta_1 = \beta_2$, $g(x)$ is even with respect to the
point $-(\mu_1 + \mu_2)/2$ while $g_H(y)$ is odd.  
Finally  note that  in the  low temperature  limit  $\beta \rightarrow
\infty$ from 
the asymptotic expression of the digamma function\cite{abramowitz:stegun}
\begin{equation}
  \label{eq:asy:psi}
  \Psi(z) = \ln(z) - \frac{1}{2 z} + O(z^{-2})
\end{equation}
the elementary result 
\begin{equation}
  \label{eq:gH:beta:inf}
  g_H(y) = \frac{1}{\pi} \ln \left| \frac{y-\mu_2}{y-\mu_1} \right| + O(\beta^{-2})
\end{equation}
is obtained.\cite{king}

The Fourier transform  of $g(x)$ can be easily  computed with a slight
modification of the above procedure. 
%
In fact  choosing $\lambda>0$ and  using the Jordan lemma  closing the
contour in $\Im (x) >0 $ one obtains 
\begin{equation}
  \label{eq:gF:2}
  \begin{split}
  g_F(\lambda) & =
  \int_{-\infty}^{+\infty} \de x \, \e^{i\, \lambda\,x} \,g(x) \\
  & =\pi i 
  \left[ \frac{\beta_2^{-1} \e^{i\,\lambda\,\mu_2}}{\sinh(\pi\lambda/\beta_2)} -
    \frac{\beta_1^{-1}  \e^{i\,\lambda\,\mu_1} }{\sinh(\pi \lambda/\beta_1)}
  \right].
\end{split}
\end{equation}
The same  expression is obtained choosing $\lambda<0$  and closing the
contour in $\Im(x)<0$. 
Notice  that from  this result  one  can easily  evaluate the  Fourier
transform of $g_H$ by mean of the general relation\cite{king}

\begin{equation}
  \label{eq:fourier:hilbert}
   \int_{-\infty}^{+\infty}\de \,x \, \mathrm{e}^{i \lambda\, x}\, g_H(x) = -i
   \, \mathrm{sign} (\lambda) \, g_F(\lambda). 
\end{equation}

\section{Evaluation of an integral }
\label{sec:int:ggH}
 
Let consider the special  case $g_H(x; \omega, -\omega, \beta, \beta)$
and $g(x;0,-V, \beta, \beta)$, the integral
\begin{equation}
  \label{eq:Ia:1:vol}
  I = \int_{-\infty}^{+\infty} \,g(x;0,-V, \beta, \beta) \; 
  g_H(x; \omega, -\omega, \beta, \beta) \de x, 
\end{equation}
describes the  asymmetrical contribution to  the current flowing  in a
molecular  wire\cite{appl:prb}.  To avoid  convergence problems  it is
better to evaluate the Fourier integral 
\begin{equation}
  \label{eq:Ia:1}
  K(\lambda) = \int_{-\infty}^{+\infty} \e^{i\,\lambda\,x}\,g(x;0,-V, \beta, \beta) \; 
  g_H(x; \omega, -\omega, \beta, \beta) \de x, 
\end{equation}
from which one obtains $I = K( \lambda \rightarrow 0)$. 
In fact again using the Jordan lemma with $\lambda >0$ one needs the 
poles structure of the integrand in $\Im(x)>0$.  
Let us  define some intermediate  quantities to simplify  the notation
$c_0 = \coth(\pi\, v_0)$, $c_{\pm} = \coth (\pi \, v_{\pm} )$, 
$ \eta_{\pm \omega} = - ( \pm c_0 +
    c_{\mp})/2
$    where   $v_0    =   \beta    \omega/2\,\pi$   and    $v_{\pm}   =
\beta(V\pm\omega)/2\pi$. 
Moreover let $z = \exp( -2\pi\lambda/\beta) <1 $.  
The poles are found
straightforwardly:
\begin{itemize}
\item[a)] $x_n =  \omega + (n+1/2)2\pi i/\beta$, $n=0,1,2,\ldots$ with
  residua
  \begin{equation}
      \label{eq:res:p1}
      r_{+\omega}(n)  
      = \frac{i}{\beta} \e^{i\, \lambda\,\omega}
      z^{n+1/2}\,\eta_{+\omega}
    \end{equation}
  \item[b)] $x_n = -\omega + (n+1/2)2\pi i/\beta$, $n=0,1,2,\ldots$ with
    residua 
    \begin{equation}
      \label{eq:res:p3}
      r_{-\omega}(n)  
      = \frac{-i}{\beta} \e^{-i\, \lambda\,\omega}
      z^{n+1/2}\,\eta_{-\omega}
    \end{equation}

  \item[c)] $x_n = (n+1/2) 2 \pi i/\beta $, $n=0,1,2,\ldots $, with residua
    \begin{equation}
      \label{eq:res:q1}
      r_0(n)  = \frac{-i}{\pi  \beta} z^{n+1/2}\,\Im\left[
        \Psi(1+n+i\,v_0)+ \Psi(-i\, v_0 -n)\right];
    \end{equation}
  \item[d)]  $x_n  = V  +  (n+1/2)2  \pi  i/\beta$, $n=0,1,2,\ldots$  with
    residua 
    \begin{equation}
      \label{eq:res:q3}
      \begin{split}
        r_V(n) & = \frac{\e^{ i\,\lambda\,V}}{2\pi\beta} z^{n+1/2}
        \times \\
        &\phantom{=} \bigg[ \Psi(i\,v_-  -n) + \Psi(- i\,v_- +  n + 1)
         \\
        &\phantom{=}    -\Psi(i\,v_+  -n)-  \Psi(-i\,v_+ +  n  +
          1)\bigg] .
      \end{split}
    \end{equation}
  \end{itemize}
The integral is thus expressed as 
\begin{equation}
  \label{eq:K:somme}  
K(\lambda) = 2 \pi i \sum_{n=0}^{+\infty} 
\left[ r_{+\omega}(n) + r_{-\omega}(n) + r_{0}(n) + r_{V}(n) \right]. 
\end{equation}
The first two terms are trivial geometric series whose result is
\begin{equation}
  \label{eq:res:p1ep3}
  \begin{split}
  R_{\omega}  & \equiv  \sum_{n=0}^{+\infty}  \left[ r_{+\omega}(n)  +
    r_{-\omega}(n) \right] \\ 
  & = 
  \frac{i}{2\,\beta}\;\frac{1}{\sinh(\pi\,\lambda/\beta)}
\left( \eta_{+\omega}\e^{i\,\lambda\,\omega} - 
  \eta_{-\omega}\e^{-i\,\lambda\,\omega}\right) .
\end{split}
\end{equation}
For      the     third     term      using     the      results     of
Appendix~\ref{sec:summation-formula} one gets
\begin{equation}
  \label{eq:R0}
  R_0  \equiv \sum_{n=0}^{+\infty} r_{0}(n) 
  = \frac{-i\,z^{1/2} }{\pi\beta}\,\left\{ \Im \left[S_+(i\,v_0,z) \right] - 
    \frac{\pi\,c_0}{1-z}\right\};
\end{equation}
similarly for the last term one finds
\begin{equation}
  \label{eq:RV}
  \begin{split}
    R_V & \equiv \sum_{n=0}^{+\infty} r_{V}(n) 
    = \frac{\e^{ i\,\lambda\,V}z^{1/2}}{\pi\beta} \times\\
    &\phantom{=}  \bigg\{ 
     S_{+}(-i\,v_-,z) - S_+(-i\,v_+, z) 
    +i \frac{\pi}{2} \frac{c_- - c_+}{1-z}
 \bigg\} .
\end{split}
\end{equation}
The integral is now completely solved as
\begin{equation}
  \label{eq:K:lamda:fin}
  K(\lambda) = 2\pi\,i \left( R_{\omega} + R_0 + R_{V} \right).
\end{equation}

To obtain the limiting value $K(0)$ one needs to carefully extract the
divergent terms from $R_{\omega}$, $R_0$ and $R_V$
\begin{align}
  \label{eq:asymp:R}
  R_{\omega} & = \frac{i\,(\eta_{+\omega} - \eta_{-\omega})}{2\pi}\,
  \frac{1}{\lambda} \nonumber \\
  & \phantom{=}  -\frac{\omega}{2\pi}(\eta_{+\omega} + \eta_{-\omega})
  + O(\lambda) \\
R_{0} +  R_V &  = - \frac{i\,(\eta_{+\omega}  - \eta_{-\omega})}{2\pi}\,
\frac{1}{\lambda} \nonumber \\
& \phantom{=} + \frac{i}{\pi\beta} \bigg[ 2\, h(v_0) + h(v_-) - h(v_+)
\bigg] \nonumber \\
& \phantom{=} - \frac{\omega}{4\pi} (c_- + c_+ ) + O(\lambda),
\end{align}
where  the property\cite{abramowitz:stegun} 
 $\Im(
\Psi(i\,x) ) = 1/(2\,x) + (\pi/2) \, \coth(\pi\,x)$ is used and the function $h(x)$ is 
\begin{equation}
  \label{eq:def:h(x)}
  h(x) \equiv x \Re \big[ \Psi(i\,x) \big].
\end{equation}

As expected the divergent terms cancel each other and one gets
the final result
\begin{equation}
  \label{eq:K0}
  I = \frac{2}{\beta}\bigg[ h(v_+) -h(v_-) - 2h(v_0) \bigg].
\end{equation}

Using the asymptotic expansion\cite{abramowitz:stegun} $h(x) = x\,\ln(x) +1/(12x) + 1/(120x^3)
+ O(x^{-5})$ 
the limit of low temperature is obtained
\begin{equation}
  \label{eq:low:temp}
  \begin{split}
  I & = \frac{1}{\pi}\bigg[ (V+\omega)\ln(V+\omega) - (V-\omega)\ln(V-\omega) -
  2\omega\ln\omega \bigg] \\ 
  & \phantom{=} -\frac{2\,\pi\, V^2}{\omega\,(V^2 -
    \omega^2)}\, \beta^{-2} + O(\beta^{-4})
\end{split}
\end{equation}
where   the   first  term   is   related   to   the  elementary   case
\eqref{eq:gH:beta:inf}.  On   the  other   hand  for  small   $x$  one
has\cite{abramowitz:stegun}  $h(x)  =  \Psi(1)\,  + \zeta(3)  \,x^3  +
O(x^5)$ ($\zeta(3) \approx 1.202056$) 
and the limit of high temperature
follows
\begin{equation}
  \label{eq:high:temp}
  I = \frac{ 12 \, \zeta(3) \, \omega \, V^2}{(2\pi)^3} \, \beta^2 +
  O(\beta^4). 
\end{equation}

\section{Conclusions}
\label{sec:concl}
The paper derives analytical results for some integrals related to the
difference of two Fermi  functions: its Hilbert transform is evaluated
in term of  the digamma function while the  Fourier transform involves
only elementary functions.  Additionally an exact result useful in the
discussion  of  the  asymmetric  part  of the  electronic  current  in
presence of phonon scattering is reported.

\appendix
 
\section{ \label{sec:summation-formula} A summation formula}
The series
\begin{equation}
  \label{eq:sum:psi}
  S_+(w, z) = \sum_{n=0}^{+\infty} \Psi(w + n + 1 ) \, z^n \qquad |z| < 1,
\end{equation}
can  be expressed  in  a  closed form\cite{serbo:digamma}. In  fact
exploiting the telescoping properties of $\Psi$
\begin{equation}
  \label{eq:Psi:tel}
  \Psi(w+n+1) = \sum_{k=0}^{n}\frac{1}{k+w} + \Psi(w)
\end{equation}
and multiplying by $z^n$ and summing one gets    
\begin{equation}
  \label{eq:S+:der1}
  S_+(w,z) = \frac{1}{1-z}\,\Psi(w) + \sum_{n=0}^{+\infty}\sum_{k=0}^{n} \frac{z^n}{k+w}. 
\end{equation}
The double sum can be decoupled using the rule 
\begin{equation}
  \label{eq:rule:dec}
  \sum_{n=0}^{+\infty}\sum_{k=0}^{n} A_{n,k} = \sum_{n=0}^{+\infty}\sum_{k=0}^{+\infty} A_{n+k,k}
\end{equation}
thus obtaining
\begin{equation}
  \label{eq:def:S+}
  S_+(w,z) = \frac{1}{1-z}\left[ \Psi(w) + \phi(z,w) \right]
\end{equation}
where   the  last   term   is   a  particular   case   of  the   Lerch
transcendent\cite{erdelyi}   $\phi(z,w) = \Phi(z,1,w) $ 
\begin{equation}
  \label{eq:Lerch}
  \Phi(z,s,w) \equiv \sum_{n=0}^{+\infty} \frac{z^n}{(n+w)^s}. 
\end{equation}
Notice  that   $\phi$  can  be   also  expressed  in  term   of  Gauss
hypergeometric function\cite{erdelyi}
\begin{equation}
  \label{eq:f:2F1}
  \phi(z,w) = \frac{1}{w}\, \phantom{}_2F_1( 1, w ; 1+w; z),  
\end{equation}
and has  a logarithmic  singularity in $z=1$  as can checked  from the
expansion\cite{abramowitz:stegun}
\begin{equation}
  \label{eq:ln(1-z)}
  \phi(z,w) = \sum_{n=0}^{\infty}\frac{(w)_n}{n!}\,(1-z)^n\,
\left[ \Psi(n+1) -\Psi(w+n) - \ln(1-z) \right]
\end{equation}
which is valid if $|1-z|<1$ and $|\arg(1-z)|<\pi$. 

From the reflection  property  of the  digamma $\Psi(1-z)=\Psi(z)  +
\pi\cot(\pi\,z)$ follows that 
$\Psi(w-n) = \Psi(w+n+1) - \pi\cot(\pi\,w)$ and thus 
\begin{equation}
  \label{eq:S-}
  \sum_{n=0}^{+\infty}  \Psi(w   -  n  )   \,  z^n  =   S_{+}(-w,z)  -
  \frac{\pi\cot(\pi\,w)}{1-z}. 
\end{equation}
Moreover
\begin{equation}
  \label{eq:S-:S+}
  \sum_{n=0}^{+\infty} 
\left[ \Psi(-w - n ) + \Psi(w + n +1) \right] z^n = 2 S_{+}(w,z) + 
  \pi \frac{\cot(\pi\,w)}{1-z}. 
\end{equation}

When  $z \rightarrow  1^{-}$  along  the real  axis,  the behavior  of
$S_{+}(w,z)$ is easily extracted from \eqref{eq:ln(1-z)}
\begin{equation}
  \label{eq:S+:asymp}
\begin{split}
S_{+}(w,z)  & =  \frac{\Psi(1) -  \ln(1-z)}{1-z} +  w  \left[\Psi(2) -
  \Psi(w+1) \right] \\
&\phantom{=} -w \ln(1-z) + O(1-z).
\end{split}
\end{equation}

\bibliographystyle{apsrev}
\bibliography{ref}

\end{document}